\documentclass[aps,pra]{revtex4-1}

\usepackage[utf8]{inputenc}
\usepackage{amsmath}
\usepackage{amssymb}
\usepackage{bm}
\usepackage{hyperref}
\usepackage{tikz}

\def\sumint{\sum\kern-15pt\int}
\def\RABITT{RABBITT}

\begin{document}

\title{Dipole-laser coupling delay in two-color (\RABITT) photoionization of polar molecules}

\author{Jakub Benda}
\email{jakub.benda@matfyz.cuni.cz}
\affiliation{Institute of Theoretical Physics, Faculty of Mathematics and Physics, Charles University, V Holešovičkách 2, Praha 8, 180 00, Czech Republic}

\author{Zdeněk Mašín}
\affiliation{Institute of Theoretical Physics, Faculty of Mathematics and Physics, Charles University, V Holešovičkách 2, Praha 8, 180 00, Czech Republic}

\date{\today}

\begin{abstract}
We study theoretically the reconstruction of attosecond beating by interference of two-photon transitions (\RABITT{}) in strongly polar molecules. The time-dependent energy of a polar molecule in the infrared (IR) field gives rise to an additional dipole-laser coupling contribution to the sideband delay. In a time-independent picture this translates to the initial state becoming a linear combination of IR-dressed states. We extend the recently developed time-independent molecular $R$-matrix method to include the additional interfering ionization pathways arising from the IR-dressed initial state and obtain very good agreement with a reference non-perturbative time-dependent \RABITT{} simulation. We discuss the asymptotic behaviour of such ionization amplitudes and recover a known approximate asymptotic formula for the dipole-laser coupling delay derived earlier in the context of attosecond streaking. At low photon energies the dipole-laser coupling contributes significantly even in an unoriented molecular sample. Finally, we show that in photodetachment of polar singly charged negative ions the sideband delay is asymptotically proportional only to Wigner delay.
\end{abstract}

\maketitle

\section{Introduction}
The experimental technique of reconstruction of attosecond beating by interference of two-photon transitions (\RABITT) has progressed from the initial applications to characterization of attosecond pulses~\cite{paul2001} to a sensitive probe of ultrafast dynamics in atoms~\cite{bharti2021,Dahlstrom,fuchs2020,joseph2020,klunder2011,isinger2017,guenot2012,cattaneo2016} and molecules~\cite{BaykushevaWorner,cattaneo2018,CF4,Hockett,KamalovCO2,Loriot2017,Nandi,haessler2009,vos2018,boutu2008,RM-RABITT}. In \RABITT{} the system of interest is photoionized by a combination of temporally overlapping extreme ultraviolet (XUV) and infrared (IR) pulses comprising two neighbouring XUV harmonics of a high harmonic source and the driving phase-locked IR field. This sets up conditions for the interference of two two-photon ionization pathways leading to the appearance of a side-band in the photoelectron spectrum that is sensitive to the relative phase of the two contributing photoionization amplitudes. Measured as a function of XUV-IR delay, the photoelectron sideband displays a beating whose offset determines the requisite phase difference. Expressed in units of time this relative phase gives access to the photoionization time delay
\begin{eqnarray}\label{eq:tau_sb}
\tau_{sb} = \frac{1}{2\omega}\mathrm{arg}(d_{+}^{(2)*}d_{-}^{(2)}),
\end{eqnarray}
where $d_{-}^{(2)}$ and $d_{+}^{(2)}$ are the two-photon matrix elements~\cite{RM-RABITT} connecting the ground state with the sideband through XUV absorption and IR emission and through XUV absorption and IR absorption, respectively.

Despite its long history, \RABITT{} has been extended to molecules only recently to study electronic~\cite{BaykushevaWorner,vos2018,Hockett,KamalovCO2,Loriot2017,boutu2008} and nuclear~\cite{CF4,cattaneo2018,Nandi,haessler2009,borras2023,gong2023} dynamics, including the dynamics in the vicinity of resonances in molecules~\cite{Nandi} and solid targets~\cite{Kasmi,Ambrosio}. This reflects the complexity of the process both from the experimental as well as from the theoretical perspective. The photoionization time delay is usually interpreted with the help of the so-called asymptotic approximation~\cite{Dahlstrom} which splits the sideband delay into two components
\begin{eqnarray}\label{eq:asymptotic_delay}
\tau_{sb} \approx \tau_{W} + \tau_{cc},
\end{eqnarray}
where $\tau_{W}$ is the intrinsic one-photon Eisenbud-Wigner-Smith time delay and $\tau_{cc}$ is the continuum-continuum coupling delay~\cite{ivanov2011} coming from the IR absorption or emission in the continuum. This approximation is applicable only at sufficiently high photoelectron energies and depends on the system~\cite{RM-RABITT}.

We have shown recently~\cite{RM-RABITT} that this asymptotic picture is not complete since the IR photon can be absorbed or emitted not only by the continuum electron but also in the residual ion. This effect requires near-resonant conditions between the ion transition and the IR laser frequency and is amplified near resonances. As a consequence Eq.~\eqref{eq:asymptotic_delay} contains an additional term $\tau_{coupl}$ caused by the laser-ion coupling.

In this paper we show that in polar molecules there is an additional delay component $\tau_{dLC}$, which reflects the dynamics of molecules dressed in the laser field. Therefore, \RABITT{} delay in a polar molecule generally comprises of four terms:
\begin{eqnarray}\label{eq:asymptotic_delay_dLC}
\tau_{sb} \approx \tau_{W} + \tau_{coupl} + \tau_{dLC} + \tau_{cc}.
\end{eqnarray}
Of course, this picture breaks down at low energies where the separability of the absorption events between the ion and the photoelectron is not possible and the system absorbs as a whole: A complex dynamics intertwining all the effects described above takes place~\cite{RM-RABITT,zhang2010}. This is inherently a multi-electron process which must be described by an appropriate level of theory.

Recently, we developed a full multi-photon stationary approach to above-threshold photoionization of multielectron molecules based on the $R$-matrix method~\cite{multiphoton} which is capable of calculating directly the two-photon matrix elements from Eq.~\eqref{eq:tau_sb} without the asymptotic approximation. Since laser-dressed states have a time-dependent energy the original stationary approach must be extended to incorporate the dressing field. We show that such laser-dressed two-photon amplitudes give time delays in perfect agreement with our explicit time-dependent calculations using $R$-matrix with time (RMT)~\cite{RMT}. Furthermore, we show that the effects of laser dressing manifest separately in the initial and the final states of the molecule. In orientation-averaged measurements, the dipole-laser coupling delay vanishes at high energies, but it persists at low energies and constitutes a strong effect in molecular-frame data. We demonstrate this effect in the LiH molecule which has a large dipole moment of $5.88$~D~\cite{nelson1967}.

In Secs.~\ref{sect:one-photon} and~\ref{sect:two-photon} we review the derivation of one- and two-photon ionization amplitudes of a non-polar molecule from time-dependent perturbation theory. In Sec.~\ref{sect:dressing} we extend the description to consider also the dynamics of the initial state of a polar molecule driven by a low-frequency electric field. The general equations are simplified in Sec.~\ref{sect:apx} for typical field strengths that are used in \RABITT{} experiments. In Sec.~\ref{sect:oavg} we discuss the effect of the averaging over molecular orientations and in Sec.~\ref{sect:asym} we derive the asymptotic formula for the dipole-laser coupling delay from the RABITT{} formalism introduced before. In Sec.~\ref{sect:results} we show the results from numerical simulations demonstrating accuracy of the time-independent theory and we discuss regions of validity of the dipole-laser coupling delay formula. We summarize in Sec.~\ref{sect:conclusion}.

\section{One-photon ionization} \label{sect:one-photon}

In this section we review standard time-dependent perturbation theory of photoionization~\cite{FaisalBook}. We work in the fixed-nuclei approximation and use Hartree atomic units unless stated otherwise. The electronic initial state of a non-polar molecule is
\begin{equation}
    \Psi_i(t) = \Psi_i \mathrm{e}^{-\mathrm{i} E_i t} \,,
\end{equation}
satisfying the time-dependent Schrödinger equation (TDSE)
\begin{equation}
    \mathrm{i} \frac{\mathrm{d}}{\mathrm{d}t} \Psi_i(t) = H_0 \Psi_i(t) \,.
\end{equation}
Here \(H_0\) is the time-independent Hamiltonian operator describing the molecular structure,
\begin{equation}
    H_0 \Psi_i = E_i \Psi_i \,.
\end{equation}
We search for the time-dependent state of the system \(\Psi(t)\) once the ionizing field \(\bm{F}(t)\) is applied. Then, in the length gauge, we solve the equation
\begin{equation}
    \mathrm{i} \frac{\mathrm{d}}{\mathrm{d}t} \Psi(t) = [H_0 - \bm{D} \cdot \bm{F}(t)] \Psi(t) \,,
    \label{eq:tdse}
\end{equation}
where \(\bm{D}\) is the electronic dipole operator. The contribution of the nuclei to the total dipole of the system is not considered, because the energy of the fixed-in-space nuclei in the ionizing field is irrelevant to the properties of the electronic wave function \(\Psi(t)\).
In the deep past, the state \(\Psi(t)\) will coincide with the initial state,
\begin{equation}
    \Psi(t) \rightarrow \Psi_i(t) \qquad [t \rightarrow -\infty] \,,
\end{equation}
which we take as the initial condition for the solution of Eq.~\eqref{eq:tdse}. Furthermore, in the spirit of perturbation theory, we declare the time-dependent term of the total Hamiltonian in Eq.~\eqref{eq:tdse} a small perturbation and say that in the zeroth order, even the field-affected solution behaves as the field-free one,
\begin{equation}
    \Psi(t) \approx \Psi_i(t) \qquad [\text{0th order of PT}] \,.
    \label{eq:0th}
\end{equation}
To get an improvement to the next order we begin by expanding the solution in the time-independent eigenstate basis:
\begin{equation}
    \Psi(t) = \sumint\limits_k c_k(t) \Psi_k \mathrm{e}^{-\mathrm{i}E_k t} \,,
    \label{eq:psit-expansion}
\end{equation}
where the continuous part corresponds to proper scattering states.
After substitution into Eq.~\eqref{eq:tdse}, elimination of terms that occur on both sides, and projection of both sides of the equation on \(\Psi_k\), we obtain
\begin{equation}
    \dot{c}_k(t) = \mathrm{i} \sumint\limits_l \langle \Psi_k | \bm{D} | \Psi_l \rangle \cdot \bm{F}(t) c_l(t) \mathrm{e}^{\mathrm{i} (E_k - E_l) t} \,.
    \label{eq:dot_ckt}
\end{equation}
Now, on the right-hand side we assume the unperturbed initial state, which is equivalent to
\begin{equation}
    c_l(t) = c_l^{(0)}(t) = \delta_{li} \,.
\end{equation}
This leads to the final equation for other coefficients \(k \ne i\),
\begin{equation}
    \dot{c}_k^{(1)}(t) = \mathrm{i} \bm{D}_{ki} \cdot \bm{F}(t) \mathrm{e}^{\mathrm{i}(E_k - E_i)t} \,.
\end{equation}
The electric field is taken as a cosine standing wave, equivalent to the superposition of two traveling waves,
\begin{equation}
    \bm{F}(t) = \bm{F}_1 \cos (\Omega t) = \tfrac{1}{2}\bm{F}_1 \mathrm{e}^{+\mathrm{i}\Omega t} + \tfrac{1}{2}\bm{F}_1 \mathrm{e}^{-\mathrm{i}\Omega t} \,.
\end{equation}
In this paper we always consider only one of the two traveling waves, corresponding to absorption or emission of a photon.
While physical fields always have a finite duration and hence also a nontrivial spectral distribution, restriction to monochromatic fields of infinite duration makes the subsequent transition to a time-independent picture more straightforward. This approximation is very common in the analysis of \RABITT{}~\cite{Dahlstrom}.

We solve the equation for \(c_k^{(1)}(t)\) by integration,
\begin{equation}
    c_k^{(1)}(t) = \mathrm{i} \bm{D}_{ki} \cdot \tfrac{1}{2}\bm{F}_1 \int\limits_{-\infty}^t \mathrm{e}^{\mathrm{i}(E_k - E_i - \Omega)\tau} \mathrm{d} \tau \,.
    \label{eq:ck1}
\end{equation}
In the limit of long times we have
\begin{align}
    c_k^{(1)}(t \rightarrow +\infty)
    &\rightarrow \mathrm{i} \bm{D}_{ki} \cdot \tfrac{1}{2}\bm{F}_1 \int\limits_{-\infty}^{+\infty} \mathrm{e}^{\mathrm{i}(E_k - E_i - \Omega)\tau} \mathrm{d} \tau \nonumber \\
    &= 2\pi\mathrm{i} \delta(E_k - E_i - \Omega) \bm{D}_{ki} \cdot \tfrac{1}{2}\bm{F}_1 \,.
\end{align}
That is, the amplitude of transition to a specific final state is proportional to the Dirac delta function that enforces energy conservation, and to the field-parallel component of the transition dipole. Here we additionally assume that in weak stationary electromagnetic fields the solution of the time-dependent Schrödinger equation also eventually reaches a stationary character, where the coefficients pertaining to individual final states no longer change. Only then can we meaningfully evaluate them in the limit of long times and transit to the time-independent picture. While this is a standard setting of the perturbation theory, there certainly exist cases featuring non-perturbative phenomena like Rabi oscillations that defy transiting to a single stationary state. Such processes are not considered in this work.

Positive-energy eigenstates \(\Psi_k\) used in Eq.~\eqref{eq:psit-expansion} can be chosen as the stationary photoionization states that in large distances converge to a product of a residual ion state \(\Phi_k\) and a photoelectron wave-function \(\psi_k\). Then the Wigner delay associated with ionization from the real initial state \(\Psi_i\) into a specific real residual state \(\Phi_k\) can be calculated from the one-photon transition dipole elements \(\bm{D}_{ki}\),
\begin{equation}
    \tau_{W,ki}(\bm{k}_k) = \frac{\mathrm{d}}{\mathrm{d}E_k} \arg \psi_{k}^*(\bm{k}_k) = \frac{\mathrm{d}}{\mathrm{d}E_k} \arg [ \bm{D}_{ki} \cdot \bm{F}_1 ] \,,
\end{equation}
where \(E_k = k_k^2/2\) is the asymptotic photoelectron kinetic energy.

\section{Two-photon ionization} \label{sect:two-photon}

As we have seen in the preceding section from the combination of Eqs.~\eqref{eq:psit-expansion} and~\eqref{eq:ck1}, after absorption of the first photon the system is in the state
\begin{equation}
    \Psi(t) = \mathrm{i} \sumint\limits_l (\bm{D}_{li} \cdot \tfrac{1}{2}\bm{F}_1) \Psi_l \mathrm{e}^{-\mathrm{i} E_l t}
    \int\limits_{-\infty}^t \mathrm{e}^{\mathrm{i}(E_l - E_i - \Omega_1)\tau} \mathrm{d} \tau \,.
\end{equation}
Once again, we will use this state as the initial condition for the next iteration of the TDSE, Eq.~\eqref{eq:dot_ckt}. We substitute \(c_k(t)\) from Eq.~\eqref{eq:ck1} for \(c_l(t)\) on the right-hand side of Eq.~\eqref{eq:dot_ckt}, arriving at
\begin{equation}
    \dot{c}_k^{(2)}(t) = -\sumint\limits_l (\bm{D}_{kl}\cdot\tfrac{1}{2}\bm{F}_2) (\bm{D}_{li}\cdot\tfrac{1}{2}\bm{F}_1)
    \mathrm{e}^{\mathrm{i} (E_k - E_l - \Omega_2) t}
    \int\limits_{-\infty}^t \mathrm{e}^{\mathrm{i}(E_l - E_i - \Omega_1)\tau} \mathrm{d} \tau \,,
\end{equation}
which leads to
\begin{equation}
    c_k(T) = -\sumint\limits_l (\bm{D}_{kl}\cdot\tfrac{1}{2}\bm{F}_2) (\bm{D}_{li}\cdot\tfrac{1}{2}\bm{F}_1)
    \int\limits_{-\infty}^{T} \mathrm{e}^{\mathrm{i} (E_k - E_l - \Omega_2) t}
    \int\limits_{-\infty}^t \mathrm{e}^{\mathrm{i}(E_l - E_i - \Omega_1)\tau} \mathrm{d} \tau \mathrm{d} t \,.
    \label{eq:ckT2}
\end{equation}
If a proper regularization factor \(\epsilon \rightarrow 0+\) is introduced to damp the integrand for \(t \rightarrow -\infty\), the inner integral can be written as
\begin{equation}
    \int\limits_{-\infty}^t \mathrm{e}^{\mathrm{i}(E_l - E_i - \Omega_1 - \mathrm{i}\epsilon)\tau} \mathrm{d} \tau
    = \frac{1}{\mathrm{i}} \frac{\mathrm{e}^{\mathrm{i}(E_l - E_i - \Omega_1)t}}{E_l - E_i - \Omega_1 - \mathrm{i}\epsilon}  \,.
\end{equation}
The resulting exponential factor combines with the other exponential factor in Eq.~\eqref{eq:ckT2} and in the limit of long times leads to the Dirac $\delta$ function:
\begin{equation}
    c_k(T \rightarrow +\infty) \rightarrow
    2\pi \delta(E_k - E_i - \Omega_1 - \Omega_2) \frac{1}{\mathrm{i}} \sumint\limits_l
    \frac{(\bm{D}_{kl}\cdot\tfrac{1}{2}\bm{F}_2) (\bm{D}_{li}\cdot\tfrac{1}{2}\bm{F}_1)}{E_i + \Omega_1 - E_l + \mathrm{i}\epsilon}
    \,.
\end{equation}
That is, the amplitude of transition to a specific state after absorption of two photons is proportional to the two-photon matrix element (identical to that in, e.g., \cite{Dahlstrom}) and to the Dirac delta function that maintains energy conservation.

\section{Ionization of polar states embedded in IR field}\label{sect:dressing}

In the first section it was assumed that the initial state is not perturbed by the field. In contrast, when there is some dressing field, even if it is of low intensity and photon energy so that it cannot ionize the molecule on its own, one needs to take this field into account when the molecule is strongly polar.

We consider a dressing IR field with time-dependent intensity
\begin{equation}
    \bm{F}_{IR}(t) = \bm{F}_{IR} \cos \omega t = -\frac{\mathrm{d}}{\mathrm{d}t} \bm{A}_{IR}(t) \,.
\end{equation}
A state with a non-zero permanent dipole \(\bm{D}_{ii}\) has the time-dependent energy \(E(t)\) in this field,
\begin{equation}
    E(t) = E_0 - \bm{D}_{ii} \cdot \bm{F}_{IR}(t) \,.
\end{equation}
The time-dependent wave function of such a state has to satisfy the Schrödinger equation
\begin{equation}
    \mathrm{i} \frac{\mathrm{d}}{\mathrm{d}t} \Psi_i(t)
    = [ H_0 - \bm{D}_{ii} \cdot \bm{F}_{IR}(t) ] \Psi_i(t) \,.
\end{equation}
This is straightforward to solve by separation of variables, resulting in
\begin{equation}\label{eq:bs_en}
    \Psi_i(t) = \Psi_i \mathrm{e}^{-\mathrm{i} E_i t - \mathrm{i} \bm{D}_{ii} \cdot \bm{A}_{IR}(t)} \,.
\end{equation}
In other words, a polar molecule receives an extra time-dependent phase factor if it is located in a field~\cite{Pazourek,Baggesen}. This phase now has to be included in all prior derivations. To be able to treat the state analytically, we use the Jacobi-Anger expansion~\cite[Eq.~(10.12.2)]{DLMF}, or~\cite{Faisal},
\begin{equation}
    \mathrm{e}^{\mathrm{i}x \sin \phi} = \sum_{m = -\infty}^\infty J_m(x) \mathrm{e}^{\mathrm{i}m \phi} \,,
\end{equation}
where \(J_m(x)\) is the regular Bessel function. Identifying \(\phi = -\omega t\) we can write
\begin{equation}
    \Psi_i(t) = \Psi_i \sum_{m = -\infty}^\infty J_m(\bm{D}_{ii} \cdot \bm{A}_{IR}) \mathrm{e}^{-\mathrm{i} (E_i + m\omega)t} \,.
    \label{eq:inispect}
\end{equation}
This transforms the task to an already solved problem. As in the previous two Sections we can now explicitly solve the full time-dependent problem of ionization of the polar molecule within perturbation theory by integrating out the time-dependence. The only difference is that the initial state is a linear combination of terms with different time-dependence of the phase factors. Consequently, all resulting time-independent transition amplitudes will also become linear combinations with modified energies of the initial state. For the one- and two-photon amplitudes we get
\begin{align}
    c_l^{(1)} &= 2\pi \sum_{m = -\infty}^\infty J_m(\bm{D}_{ii}\cdot \bm{A}_{IR}) \delta(E_l - E_i - \Omega_1 - m\omega) \underbrace{\mathrm{i} \bm{D}_{li}\cdot\tfrac{1}{2}\bm{F}_1}_{T_{li}^{(1)}(E_i + m\omega;\,\Omega_1)} \,, \label{eq:ck1dressed} \\
    c_k^{(2)}
    &= 2\pi \sum_{m = -\infty}^\infty J_m(\bm{D}_{ii}\cdot\bm{A}_{IR}) \delta(E_k - E_i - \Omega_1 - \Omega_2 - m\omega) \nonumber \\
    &\qquad\qquad\qquad \times \underbrace{\frac{1}{\mathrm{i}} \sumint\limits_l
    \frac{(\bm{D}_{kl}\cdot\tfrac{1}{2}\bm{F}_2)(\bm{D}_{li}\cdot\tfrac{1}{2}\bm{F}_1)}{E_i + \Omega_1 + m\omega - E_l + \mathrm{i}\epsilon}}_{T_{ki}^{(2)}(E_i + m\omega;\,\Omega_1,\,\Omega_2)} \,.
    \label{eq:ck2dressed}
\end{align}
We have obtained a linear combination of possible ionization pathways, which differ in the number of IR photons (\(m\)) absorbed before the XUV absorption takes place. The initial state of the molecule no longer has a single well-defined stationary energy. Instead, in the time-independent picture, it is in a superposition of states with different energies, spaced by the IR quantum \(\omega\). This linear combination is fixed, but the relative phases of the coefficients bear all temporal information about the dynamics of the initial state in the periodic field.

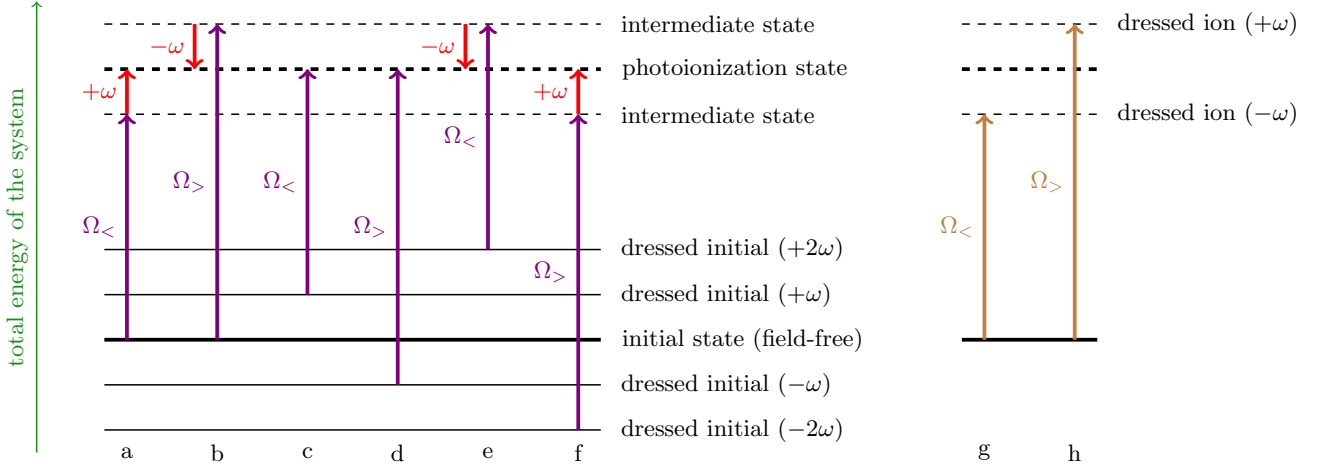
\begin{figure}[htbp]
    \centering
    \begin{tikzpicture}[scale=0.6]
        \draw[line width=0.50mm] (-0.5, 0) -- (10.5, 0) ++(0.25,0) node[right] {initial state (field-free)};
        \draw[line width=0.20mm] (-0.5, 1) -- (10.5, 1) ++(0.25,0) node[right] {dressed initial ($+\omega$)};
        \draw[line width=0.20mm] (-0.5,-1) -- (10.5,-1) ++(0.25,0) node[right] {dressed initial ($-\omega$)};
        \draw[line width=0.20mm] (-0.5, 2) -- (10.5, 2) ++(0.25,0) node[right] {dressed initial ($+2\omega$)};
        \draw[line width=0.20mm] (-0.5,-2) -- (10.5,-2) ++(0.25,0) node[right] {dressed initial ($-2\omega$)};
        \draw[line width=0.20mm, dashed] (-0.5,5) -- (10.5,5) ++(0.25,0) node[right] {intermediate state};
        \draw[line width=0.50mm, dashed] (-0.5,6) -- (10.5,6) ++(0.25,0) node[right] {photoionization state};
        \draw[line width=0.20mm, dashed] (-0.5,7) -- (10.5,7) ++(0.25,0) node[right] {intermediate state};
        \draw[->, line width=0.50mm, violet] (0, 0) -- node[left]  {$\Omega_<$} (0, 5);
        \draw[->, line width=0.50mm, red   ] (0, 5) -- node[left]  {$+\omega$} (0, 6);
        \draw[->, line width=0.50mm, violet] (2.0, 0) -- node[left] {$\Omega_>$} (2.0, 7);
        \draw[->, line width=0.50mm, red   ] (1.5, 7) -- node[left]  {$-\omega$} (1.5, 6);
        \draw[->, line width=0.50mm, violet] (6,-1) -- node[left]  {$\Omega_>$} (6, 6);
        \draw[->, line width=0.50mm, violet] (4, 1) -- node[left]  {$\Omega_<$} (4, 6);
        \draw[->, line width=0.50mm, violet] (8.0, 2) -- node[left]  {$\Omega_<$} (8.0, 7);
        \draw[->, line width=0.50mm, red   ] (7.5, 7) -- node[left]  {$-\omega$} (7.5, 6);
        \draw[->, line width=0.50mm, violet] (10,-2) -- node[left]  {$\Omega_>$} (10, 5);
        \draw[->, line width=0.50mm, red   ] (10, 5) -- node[left]  {$+\omega$} (10, 6);
        \draw[line width=0.50mm] (18.5, 0) -- (21.5, 0) ++(0.25,0) node[right] {};
        \draw[line width=0.20mm, dashed] (18.5,5) -- (21.5,5) ++(0.25,0) node[right] {dressed ion (\(-\omega\))};
        \draw[line width=0.50mm, dashed] (18.5,6) -- (21.5,6) ++(0.25,0) node[right] {};
        \draw[line width=0.20mm, dashed] (18.5,7) -- (21.5,7) ++(0.25,0) node[right] {dressed ion (\(+\omega\))};
        \draw[->, line width=0.50mm, brown] (19, 0) -- node[left]  {$\Omega_<$} (19, 5);
        \draw[->, line width=0.50mm, brown] (21, 0) -- node[left]  {$\Omega_>$} (21, 7);
        \node(a) at (0, -2.5) {a};
        \node(b) at (2, -2.5) {b};
        \node(c) at (4, -2.5) {c};
        \node(d) at (6, -2.5) {d};
        \node(e) at (8, -2.5) {e};
        \node(f) at (10, -2.5) {f};
        \node(g) at (19, -2.5) {g};
        \node(h) at (21, -2.5) {h};
        \draw[->, line width=0.20mm, green!50!black!90] (-2, -2.5) -- node[left, rotate=90, anchor=south] {total energy of the system} (-2, 7.5);
    \end{tikzpicture}
    \caption{All one-photon (odd-parity) and two-photon (even-parity) ionization pathways in three-colour ionization of a polar molecule into a state with a given photoelectron energy. In a typical \RABITT{} setup with many XUV harmonics there would be further combinations, though not significantly contributing. (a)--(f) Ionization pathways into a given stationary state of the residual ion. (g)--(h) One-photon ionization pathways into singly dressed states of the residual ion.}
    \label{fig:MPI}
\end{figure}

Pathways a--f in Fig.~\ref{fig:MPI} illustrate the possible one- and two-photon pathways that conserve energy in three-colour ionization, in a similar way as in the higher-order \RABITT{} analysis~\cite{Bertolino21,Lucchini23}.
In the three-colour ionization, the amplitude of ionization into a given final state \(k\) consists of six pathways:
\begin{align}
    T_{ki}
    &= J_0(\bm{D}_{ii}\cdot\bm{A}_{IR}) \left[ T_{ki}^{(2)}(E_i; \Omega_<, +\omega) + T_{ki}^{(2)}(E_i; \Omega_>, -\omega) \right] \nonumber \\
    &+ J_1(\bm{D}_{ii}\cdot\bm{A}_{IR}) \left[ T_{ki}^{(1)}(E_i + \omega; \Omega_<) - T_{ki}^{(1)}(E_i - \omega; \Omega_>) \right] \nonumber \\
    &+ J_2(\bm{D}_{ii}\cdot\bm{A}_{IR}) \left[ T_{ki}^{(2)}(E_i + 2\omega; \Omega_<, -\omega) + T_{ki}^{(2)}(E_i - 2\omega; \Omega_>, +\omega) \right] \,.
\end{align}
We take advantage of the relation \(J_{-m}(x) = (-1)^m J_m(x)\) valid for the integer \(m\). When the phase of the IR field is delayed by \(\omega\tau\), the \(m\)-th spectral component of the initial state~\eqref{eq:inispect} will receive an extra phase \(\exp \mathrm{i}m\omega\tau\) and the total amplitude will change to
\begin{align}
    T_{ki}(\omega\tau)
    &= J_0(\bm{D}_{ii}\cdot\bm{A}_{IR})
    \left[
        \mathrm{e}^{\mathrm{i}\omega\tau} T_{ki}^{(2)}(E_i; \Omega_<, +\omega) +
        \mathrm{e}^{-\mathrm{i}\omega\tau} T_{ki}^{(2)}(E_i; \Omega_>, -\omega)
    \right] \nonumber \\
    &+ J_1(\bm{D}_{ii}\cdot\bm{A}_{IR}) \left[
        \mathrm{e}^{\mathrm{i}\omega\tau} T_{ki}^{(1)}(E_i + \omega; \Omega_<) -
        \mathrm{e}^{-\mathrm{i}\omega\tau} T_{ki}^{(1)}(E_i - \omega; \Omega_>)
    \right] \nonumber \\
    &+ J_2(\bm{D}_{ii}\cdot\bm{A}_{IR}) \left[
        \mathrm{e}^{\mathrm{i}\omega\tau} T_{ki}^{(2)}(E_i + 2\omega; \Omega_<, -\omega) +
        \mathrm{e}^{-\mathrm{i}\omega\tau} T_{ki}^{(2)}(E_i - 2\omega; \Omega_>, +\omega)
    \right] \,.
\end{align}
Alternatively, we can write the ionization amplitude as
\begin{equation}
    T_{ki} = T_{ki,+} \mathrm{e}^{\mathrm{i} \omega \tau} + T_{ki,-} \mathrm{e}^{-\mathrm{i}\omega\tau} \,,
\end{equation}
where we collect the terms with equal \(\omega\tau\)-dependent phase factor,
\begin{align}
    T_{ki,+} &= J_0 T_{ki}^{(2)}(E_i; \Omega_<, +\omega) + J_1 T_{ki}^{(1)}(E_i + \omega; \Omega_<) + J_2 T_{ki}^{(2)}(E_i + 2\omega; \Omega_<, -\omega) \,, \\
    T_{ki,-} &= J_0 T_{ki}^{(2)}(E_i; \Omega_>, -\omega) - J_1 T_{ki}^{(1)}(E_i - \omega; \Omega_>) + J_2 T_{ki}^{(2)}(E_i - 2\omega; \Omega_>, +\omega) \,.
\end{align}
The differential ionization rate is proportional to the squared modulus of the ionization amplitude:
\begin{align}
    \frac{\mathrm{d}\sigma_{ki}}{\mathrm{d}\Omega}(\omega\tau)
    \sim |T_{ki}(\omega\tau)|^2 &= |T_{ki,+}\mathrm{e}^{\mathrm{i}\omega\tau} + T_{ki,-}\mathrm{e}^{-\mathrm{i}\omega\tau}|^2 \nonumber \\
    &= |T_{ki,+}|^2 + |T_{ki,-}|^2 + 2\mathrm{Re}\left[T_{ki,+}^* T_{ki,-} \mathrm{e}^{-2\mathrm{i}\omega\tau}\right] \nonumber \\
    &= |T_{ki,+}|^2 + |T_{ki,-}|^2 + 2|T_{ki,+}| |T_{ki,-}| \cos (2\omega\tau - \arg T_{ki,+}^* T_{ki,-}) \,.
\end{align}
Hence the experimentally observed phase shift
\begin{equation}
    2\omega\tau_2 = \arg T_{ki,+}^* T_{ki,-}
    \label{eq:\RABITT{}_tau}
\end{equation}
is a rather complicated function of one- and two-photon amplitudes. In a real experiment with many harmonics, there are also additional combinations contributing other than the six shown in Fig.~\ref{fig:MPI}.

\section{Approximations} \label{sect:apx}

Strongly polar molecules have the magnitude of the permanent dipole \(|\bm{D}_{ii}|\) on the order of 1 a.u. The vector potential of the IR field in \RABITT{} experiments is typically of the order of 0.01 a.u.\ (corresponding to a peak intensity of \(10^{10}~\text{W/cm}^2\)). Therefore, the scalar product that appears in the argument of the Bessel functions is very small and the Bessel factors can be replaced by their asymptotics:
\begin{equation}
    J_m(\bm{D}_{ii}\cdot \bm{A}_{IR}) \approx \left(\frac{\bm{D}_{ii}\cdot \bm{A}_{IR}}{2}\right)^m \frac{1}{m!}
    \qquad [m \ge 0] \,,
\end{equation}
that is, the factor \(J_0\) can be approximated by 1 and \(J_1\) by a linear term. Higher orders seem to be quite strongly suppressed. This simplifies our amplitudes to
\begin{align}
    T_{fi,+} &\approx T_{fi}^{(2)}(E_i; \Omega_<, +\omega) + \frac{1}{2} (\bm{D}_{ii}\cdot \bm{A}_{IR}) T_{fi}^{(1)}(E_i + \omega; \Omega_<) \,, \label{eq:Mp} \\
    T_{fi,-} &\approx T_{fi}^{(2)}(E_i; \Omega_>, -\omega) - \frac{1}{2} (\bm{D}_{ii}\cdot \bm{A}_{IR}) T_{fi}^{(1)}(E_i - \omega; \Omega_>) \,. \label{eq:Mm}
\end{align}
The one-photon amplitudes \(T_{fi}^{(1)}(E_i + \omega; \Omega_<)\) and \(T_{fi}^{(1)}(E_i - \omega; \Omega_>)\) are equal, because they are one-photon matrix elements of the dipole operator between the same final state and the same initial state. This is supported by the following reasoning: First, the final stationary photoionization state is given by the total energy of the system, which is equal for both amplitudes: \(E_i + \omega + \Omega_< = E_i - \omega + \Omega_> = k_f^2/2 + V_{fi}\), where \(V_{fi}\) is the ionization potential and \(k_f\) is the momentum of the photoelectron. Second, the initial states in these two amplitudes differ by energy only due to a different energy contribution of the permanent dipole to the total energy in the time-independent picture (different Floquet level); however, electronically the states are considered unchanged, see Eq.~\eqref{eq:bs_en}. In other words, for the present analysis we assume that the weak periodic field acting on the molecule in the initial state does not itself cause polarization of the target or other distortions of the electronic eigenstates. This is a reasonable approximation for non-degenerate initial states in non-resonant field conditions. Effects such as ac Stark shifts and splitting are neglected and represent additional physical phenomena acting in the case of some molecules and field configurations on top of the physics discussed here. This reasoning allows us to write both one-photon matrix elements in terms of the same transition dipole as
\begin{equation}
    T_{fi}^{(1)}(E_i + \omega; \Omega_<) = T_{fi}^{(1)}(E_i - \omega; \Omega_>) = \mathrm{i} \bm{d}_{fi}^{(1)}(k_f) \cdot \tfrac{1}{2}\bm{F}_1 \,.
\end{equation}
Note that, for brevity, the here-defined time-independent transition amplitudes \(T\) include also the amplitudes of the electric fields, which is important to remember when comparing the relative magnitudes of \(T^{(1)}\) and \(T^{(2)}\). Alternatively, one can work with scaled quantities
\begin{align}
    M_{fi,+} = \frac{\mathrm{i}}{\tfrac{1}{2}F_1 \tfrac{1}{2}F_2} T_{fi,+} &\approx \frac{\mathrm{i}}{\tfrac{1}{2}F_1 \tfrac{1}{2}F_2} T_{fi}^{(2)}(E_i; \Omega_<, +\omega) - \frac{D_{ii}}{\omega} d_{fi}^{(1)}(k_f) \,, \label{eq:Mtp} \\
    M_{fi,-} = \frac{\mathrm{i}}{\tfrac{1}{2}F_1 \tfrac{1}{2}F_2} T_{fi,-} &\approx \frac{\mathrm{i}}{\tfrac{1}{2}F_1 \tfrac{1}{2}F_2} T_{fi}^{(2)}(E_i; \Omega_>, -\omega) + \frac{D_{ii}}{\omega} d_{fi}^{(1)}(k_f) \,, \label{eq:Mtm}
\end{align}
where \(D_{ii}\) is the projection of \(\bm{D}_{ii}\) along \(\bm{A}_{IR}\), and \(d_{fi}^{(1)}\) is the projection of \(\bm{d}_{fi}^{(1)}\) along \(\bm{F}_1\).

\section{Orientation averaging} \label{sect:oavg}

From the previous derivation we obtain for the overall sideband delay, the formula:
\begin{equation}
    \tau \approx \frac{1}{2\omega} \arg \underbrace{\left[\left(M_{fi,+}^{(2)*} - \frac{D_{ii} d_{fi}^{(1)*}(k)}{\omega}\right)\left(M_{fi,-}^{(2)} + \frac{D_{ii} d_{fi}^{(1)}(k)}{\omega}\right)\right]}_{Q} \,,
    \label{eq:taua}
\end{equation}
where we abbreviated
\begin{equation}
    M_{fi,\pm}^{(2)} = \frac{\mathrm{i}}{\tfrac{1}{2}F_1 \tfrac{1}{2}F_2} T_{fi}^{(2)}(E_i; \Omega_\lessgtr, \pm\omega) \,.
\end{equation}
To get orientation-averaged results in the case of linear polarization, this has to be integrated over photoelectron emission directions \(\hat{\bm{k}}\) and averaged over polarization orientations \(\hat{\bm{\epsilon}} = \hat{\bm{\epsilon}}_{IR} = \hat{\bm{\epsilon}}_{XUV}\):
\begin{align}
    \tau_{\text{avg}} &= \frac{1}{2\omega} \arg \left[ \frac{1}{4\pi} \int Q \, \mathrm{d}\hat{\bm{k}} \,\mathrm{d}\hat{\bm{\epsilon}} \right] \nonumber \\
    &= \frac{1}{2\omega} \arg \left[\sum_{lmabcd} \left(M_{fi,+,lm,ab}^{(2)*} - \frac{D_{ii,a} d_{fi,lm,b}^{(1)*}(k)}{\omega}\right)\left(M_{fi,-,lm,cd}^{(2)} + \frac{D_{ii,c} d_{fi,lm,d}^{(1)}(k)}{\omega}\right)A_{abcd}\right],
\end{align}
where~\cite[Eq.~(6.106)]{Zamastil}
\begin{align}
    A_{abcd} = \frac{1}{15}(\delta_{ab}\delta_{cd} + \delta_{ac}\delta_{bd} + \delta_{ad}\delta_{bc}) \,.
\end{align}
The expression for the averaged time delay cannot be simplified further and the one-photon terms do not vanish for general fields. This implies that, in general, orientation averaging does not fully remove the influence of dipole-laser coupling.

\section{Asymptotic behaviour} \label{sect:asym}

In the high-energy limit, it is customary to isolate the states of the residual ion and of the photoelectron and write the wavefunction of the whole system as a product of these two parts. This separation neglects the correlation between the two sub-systems that is naturally included in the complete two-photon matrix elements presented above. Instead, in the asymptotic approximation, the effects originating in the electron-ion correlation need to be accounted for explicitly, where important.

In this particular case we need to worry about additional absorption pathways during the \RABITT{} process. These originate in laser dressing of the final ionic states. Because the residual ion is going to be dressed in very much the same way as the initial state, the first single-photon ionization will generally populate a superposition of laser-dressed ionic states with energies \(\epsilon_f + n\omega\), where \(n \in \mathbb{Z}\) and  \(\epsilon_f\) is the field-free ionic energy. While the dressing field is on, ionization into the final states that couple a given residual cation with field-free energy \(\epsilon_f\) and photoelectron with a given asymptotic kinetic energy thus may require additional absorption or emission of photons to conserve energy. Once the dressing field vanishes, these dressed states lose their field-driven energy shift and become indistinguishable from ordinary field-free states and the associated photoelectrons will interfere with the standard pathways a--f of Fig.~\ref{fig:MPI}. The two simplest ionization pathways into dressed levels, which actually consist of only one absorption each, are shown as pathways g and h in Fig.~\ref{fig:MPI}. Amplitudes for these two paths will be, again, proportional to the Bessel factors \(J_{\pm1}(\bm{D}_{ff}\cdot\bm{A}_{IR})\), only this time featuring the permanent electronic dipole moment of the residual cation. The signs in front of the Bessel functions will be opposite though, because now the dressing occurs in the bra vector and the corresponding exponential is complex conjugated. As before, we consider only the simplest pathways, because those requiring multiple transitions are suppressed. The effective single-channel amplitudes are then
\begin{align}
    T_{fi,+} &\approx \tilde{T}_{fi}^{(2)}(E_i; \Omega_<, +\omega)
        + \frac{1}{2} (\bm{D}_{ii}\cdot \bm{A}_{IR}) T_{fi}^{(1)}(E_i + \omega; \Omega_<)
        - \frac{1}{2} (\bm{D}_{ff}\cdot \bm{A}_{IR}) T_{f-\omega,i}^{(1)}(E_i; \Omega_<) \,, \label{eq:Mp-nocor} \\
    T_{fi,-} &\approx \tilde{T}_{fi}^{(2)}(E_i; \Omega_>, -\omega)
        - \frac{1}{2} (\bm{D}_{ii}\cdot \bm{A}_{IR}) T_{fi}^{(1)}(E_i - \omega; \Omega_>)
        + \frac{1}{2} (\bm{D}_{ff}\cdot \bm{A}_{IR}) T_{f+\omega,i}^{(1)}(E_i; \Omega_>) \,. \label{eq:Mm-nocor}
\end{align}
The difference of Eqs.~\eqref{eq:Mp-nocor} and \eqref{eq:Mm-nocor} from Eqs.~\eqref{eq:Mp} and \eqref{eq:Mm} is the absence of correlation between the residual ion and the photoelectron in \(\tilde{T}_{fi}^{(2)}\). In these new formulas the energy sharing between the residual ion and the photoelectron is not included. Only after this step can we proceed with the asymptotic approximation of \(\tilde{T}_{fi}^{(2)}\). Because the first-order amplitude \(T^{(1)}\) depends only on the kinetic energy of the photoelectron and not on the energy of the residual cation, the transition elements for the corresponding pathways are the same:
\begin{align}
    T_{fi}^{(1)}(E_i + \omega; \Omega_<) &= T_{f-\omega,i}^{(1)}(E_i; \Omega_<) \,, \label{eq:Tfi1}\\
    T_{fi}^{(1)}(E_i - \omega; \Omega_>) &= T_{f+\omega,i}^{(1)}(E_i; \Omega_>) \,. \label{eq:Tfi2}
\end{align}
Equation~\eqref{eq:Tfi1} corresponds to processes c and g in Fig.~\ref{fig:MPI} and the final photoelectron kinetic energy \(k^2/2 = E_i + \omega + \Omega_< - \epsilon_f\), while Eq.~\eqref{eq:Tfi2} corresponds to processes d and h and the energy \(k^2/2 = E_i - \omega + \Omega_> - \epsilon_f\). As a consequence, the interaction of the residual cation with the IR field effectively modifies the permanent electronic dipole of the initial state by the permanent electronic dipole of the final state, \(\bm{D}_{ii} \rightarrow \bm{D}_{ii} - \bm{D}_{ff}\). In the following we disregard \(\bm{D}_{ff}\), as it can be reintroduced by simple substitution into the final result.

We use the asymptotic expression for the two-photon matrix element for emission in the molecular frame~\cite[Eq.~(C8)]{RM-RABITT},
\begin{equation}
    \tilde{M}_{fi,\pm}^{(2)}
    = \frac{\mathrm{i}}{\tfrac{1}{2}F_1 \tfrac{1}{2}F_2} \tilde{T}_{fi,\pm}^{(2)}
    \approx - \mathrm{i} (\hat{\bm{k}}\cdot \hat{\bm{\epsilon}}_{IR}) A_{\kappa_\pm k} d_{fi}^{(1)}(\kappa_\pm) \,,
\end{equation}
where
\begin{equation}
    A_{\kappa k} = \frac{\mathrm{e}^{-\pi/2\kappa + \pi/2k}}{\sqrt{\kappa k}(\kappa - k)^2} \frac{(2\kappa)^{\mathrm{i}/\kappa}}{(2k)^{\mathrm{i}/k}} \frac{\Gamma(2 + \mathrm{i}/\kappa - \mathrm{i}/k)}{(\kappa - k)^{\mathrm{i}/\kappa - \mathrm{i}/k}} \,.
\end{equation}

The intermediate momentum \(\kappa_\pm\) is tied to the final momentum \(k\) by means of the energy conservation \(k^2/2 = \kappa_\pm^2/2 \pm \omega\). In the limit \(k \rightarrow +\infty\), we can then take advantage of \(k - \kappa_\pm \simeq \pm\omega/k\) and \(1/\kappa_\pm - 1/k \simeq \pm\omega/k^3\). In the end, the only contributing factors in \(A_{\kappa k}\) are the real-valued ones:
\begin{equation}
    A_{\kappa k} \rightarrow \frac{k}{\omega^2} \qquad [k \rightarrow +\infty] \,.
\end{equation}
If we further neglect the energy dependence of \(d_{fi}(k)\) in the range \(k \in (\kappa_+,\kappa_-)\), assuming \(d_{fi}^{(1)}(k) = d_{fi}^{(1)}(\kappa_+) = d_{fi}^{(1)}(\kappa_-)\), this allows us to rewrite the formula for \(Q\), cf. Eq.~\eqref{eq:taua}, as
\begin{equation}
    Q \approx \tilde{M}_{fi,+}^{(2)*}\tilde{M}_{fi,-}^{(2)} \left( 1 - \frac{D_{ii} d_{fi}^{(1)*}(\kappa_+)}{\omega \tilde{M}_{fi,+}^{(2)*}} \right) \left( 1 + \frac{D_{ii} d_{fi}^{(1)}(\kappa_-)}{\omega \tilde{M}_{fi,-}^{(2)}} \right) = \tilde{M}_{fi,+}^{(2)*}\tilde{M}_{fi,-}^{(2)} \left( 1 + \mathrm{i} \frac{D_{ii}\omega}{k\hat{\bm{k}}\cdot\hat{\bm{\epsilon}}_{IR}} \right)^2 \,.
\end{equation}
The time delay is then
\begin{equation}
    \tau \approx \frac{1}{2\omega} \arg \tilde{M}_{fi,+}^{(2)*}\tilde{M}_{fi,-}^{(2)} + \frac{1}{2\omega} \arg \left(1 + \mathrm{i} \frac{D_{ii}\omega}{\bm{k}\cdot\hat{\bm{\epsilon}}_{IR}}\right)^2 \,.
\end{equation}
The second term, corresponding to the dipole-laser coupling, can be also written as
\begin{equation}
    \tau_{\text{dLC}}^{(i)} = \frac{1}{2\omega} \tan^{-1}\, \frac{\frac{2D_{ii}\omega}{\bm{k}\cdot\hat{\bm{\epsilon}}_{IR}}}{1 - \frac{D_{ii}^2\omega^2}{(\bm{k}\cdot\hat{\bm{\epsilon}}_{IR})^2}} \,.
\end{equation}
In the last step we use the trigonometric identity
\begin{equation}
    2\tan^{-1} x = \tan^{-1} \frac{2x}{1-x^2}
\end{equation}
to obtain the final result
\begin{equation}
    \tau_{\text{dLC}}^{(i)}
    = \frac{1}{\omega} \tan^{-1}\, \frac{D_{ii}\omega}{\bm{k}\cdot\hat{\bm{\epsilon}}_{IR}}
    = \frac{1}{\omega} \tan^{-1}\, \frac{\bm{D}_{ii} \cdot \hat{\bm{\epsilon}}_{IR}}{\bm{k}\cdot\hat{\bm{\epsilon}}_{IR}} \omega
    \,.
    \label{eq:taudlc}
\end{equation}
If we now, following the earlier discussion, replace \(\bm{D}_{ii}\) by \(\Delta \bm{\mu} = \bm{D}_{ii} - \bm{D}_{ff}\), we get exactly Eq.~(6) from~\cite{Baggesen}. Taking advantage of the smallness of the argument of the arctangent to linearize the formula and split the trigonometric function in two, we can identify the additional dipole-laser coupling delay in the final state. Quoting from~\cite{Pazourek}, ``this additional time shift \(\tau_{dLC}^{(f)}\) is a true electron-electron interaction contribution absent on the [single active electron] or mean-field level.''

\section{Discussion} \label{sect:results}

\subsection{Dipole-laser coupling in numerical experiments}

\begin{figure*}[htbp]
    \centering
    \includegraphics[width=0.98\textwidth]{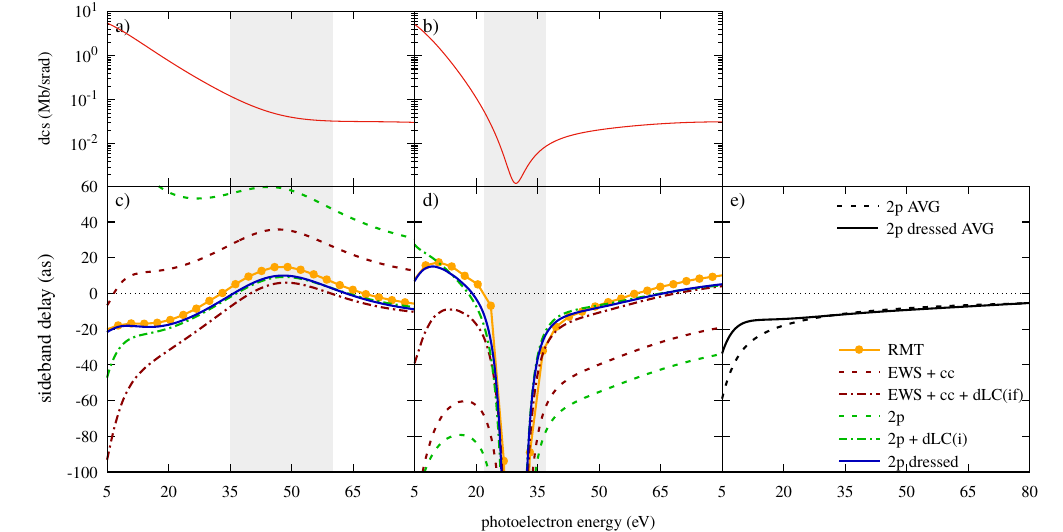}
    \caption{(a,b) Differential cross sections and (c,d,e) \RABITT{} sideband delays for photoionization of LiH from the HOMO (\(2\sigma\)) orbital in molecular frame along the molecular axis calculated in the static exchange (Hartree-Fock) model. The direction of XUV and IR is parallel to the molecular axis. (a,c) Photoemission from the lithium end. (b,d) Photoemission from the hydrogen end. (e) Emission-integrated and orientation-averaged delays -- dressed vs. undressed.}
    \label{fig:LiH}
\end{figure*}

We illustrate the success of the time-independent description in Fig.~\ref{fig:LiH}. It presents a model $R$-matrix (UKRMol+~\cite{multiphoton,masin2020}) calculation of \RABITT{} sideband delays in the LiH molecule for IR photon energy \(\omega = 0.0584\)~a.u.\ (\(\lambda = 780\)~nm). We pick this particular molecule due to its large permanent dipole moment to make the effect of the dipole-laser coupling sizable so that its contribution really stands out among the individual asymptotic terms of the \RABITT{} delay. At the same time we choose the electronically simple static-exchange model to reduce the complexity arising from many electronic channels and autoionizing resonances, even though correlated calculations are definitely feasible~\cite{LiH-escat}. This allows us to discuss dipole-laser coupling effects unobscured by other physical effects superimposed on the fundamental results.

The one-electron basis set used in the calculation comprised of the cc-pVTZ Gaussian basis centered at the nuclei and a radial center-of-mass-centered basis with partial wave expansion up to the angular momentum \(\ell = 10\) built from 30 $B$-splines of order 6 between the center of mass and the $R$-matrix radius \(R_a = 30 a_0\). The reader is referred to Ref.~\cite{masin2020} for details of the molecular $R$-matrix computational method. The calculated electronic permanent dipole moment (i.e., excluding the contribution of the nuclei) is \(D_{ii} = -3.87\)~a.u..  The electronic permanent dipole moment of the final state is \(D_{ff} = -1.54\)~a.u.

For a reference we simulated the same process using the general time-dependent ``$R$-matrix with time'' method~\cite{RMT}, which, not limited to perturbative processes, inherently considers all photon absorption orders as well as the dynamics of the initial and final molecular states in the IR field. In these calculations we used IR peak flux density \(I = 10^{11}\)~W/cm\(^2\). The corresponding peak IR field amplitudes are \(F_2 = 0.0017\)~a.u. and \(A_{IR} = 0.029\)~a.u. The inner-region model used on input to the RMT was the static-exchange model from the time-independent calculation. This guarantees the same level of description of electron correlation in both types of calculations. Therefore any discrepancy between the stationary and RMT results can be attributed only to field-dependent effects, i.e. higher-order or non-perturbative processes. Note that this is the only time-dependent calculation performed in this work; all others are time-independent.

The results in Fig.~\ref{fig:LiH} labeled ``2$p$ dressed'' are calculated from the combined amplitudes given in Eqs.~\eqref{eq:Mtp} and \eqref{eq:Mtm} using the relation~\eqref{eq:\RABITT{}_tau} and they reproduce the time-dependent RMT calculation almost perfectly. When the undressed time-independent calculation is corrected by the asymptotic expression for \(\tau_{\text{dLC}}\), Eq.~\eqref{eq:taudlc}, labeled as ``2$p$ + dLC(i)'', the agreement is also good at photoelectron energies beyond 20~eV, but below this energy the asymptotic approximation~\eqref{eq:taudlc}  separating the dipole-laser coupling delay is apparently invalid.  At higher energies, above 30~eV, the orientation averaging eliminates any effect of the dressing, as can be seen from the overlap of the ``2p AVG'' and ``2p dressed AVG'' curves. At low energies, though, some effect prevails.

Figure~\ref{fig:LiH} also presents one-photon results, that is, Eisenbud-Wigner-Smith delay corrected by the continuum-continuum delay. In this case, to achieve agreement with the time-dependent theory we need to add also the final-state dipole-laser coupling. While the second-order theory automatically takes into account the dynamics of the final state in the IR field, the one-photon method (asymptotic approximation separating the Coulomb-laser coupling delay) lacks this information and requires manual correction. Because the initial- and final-state corrections go in opposite directions, due to \(\Delta \bm{\mu} = \bm{D}_{ii} - \bm{D}_{ff}\), the deviation of the one-photon calculation from the reference time-dependent one is actually smaller than in the case of the two-photon result, which lacks only one of the two contributions.

The hill-like structure in the delays associated with emission from the Li end between roughly 35 and 65~eV (grey area in Fig.~\ref{fig:LiH}c) corresponds to a change of trend in the differential cross section in this direction (grey area in Fig.~\ref{fig:LiH}a). Similarly, the strong dip in delays associated with emission from the H end of the molecule between 20 and 35~eV (grey area in Fig.~\ref{fig:LiH}d) corresponds to a dip in the differential cross section (grey area in Fig.~\ref{fig:LiH}b). Similar structures have been investigated earlier in H$_2$ where they are caused by destructive interference between simultaneous emission from the two centers~\cite{h2plus-delays}, which results in trapping of electrons with specific angular momentum~\cite{SerovKheifets2023}, and can be described in terms of interference of distinct partial waves~\cite{RM-RABITT}.

\subsection{Photodetachment from polar negative ion molecules}

Attosecond delays in photoionization of neutral molecules are typically strongly affected by the continuum-continuum delays, arising in absorption of IR by the photoelectron coupled to the residual ion by means of the Coulomb field. In contrast, delays in photodetachment of singly-charged anions are unaffected by continuum-continuum delays due to the absence of the strong monopole interaction between the residual ion and the photoelectron. This allows for the possibility to unearth more subtle delay contributions stemming from further terms of the multipole expansion of potential experienced by the photoelectron. In particular we investigate the possibility of observing a delay caused by absorption of photon by the photoelectron while it is moving in the dipolar (rather than Coulomb) potential of the residual ion. We show that in the high-energy limit this delay contribution is imperceptible.

While here we are primarily interested in neutral residual molecules, the discussion below is for the general case of a charged residual molecular ion, i.e. for the photoelectron moving in the combination of dipolar and Coulomb fields.

In all stages of the asymptotic approach to attosecond delays as presented in~\cite{RM-RABITT}, the molecular states are eventually projected on the stationary photoionization states with the asymptotic form~\cite{Burke}
\begin{equation}
    \Psi_{f\bm{k}}^{(-)} = \sum_{lm} \mathrm{i}^{-l} \mathrm{e}^{\mathrm{i}\sigma_l}
    Y_{lm}^{*}(\bm{k}) \sum_{npk} F_{npk,flm}^{(-)}(r) Y_{pk}(\bm{r}) \Phi_n \,.
    \label{eq:photo_wf}
\end{equation}
Here the radial part is proportional to
\begin{equation}
    F_{npk,flm}^{(-)}(r) \sim H_{lm}^+(r) \delta_n^f \delta_p^l \delta_k^m - O(H_{lm}^-) \,,
    \label{eq:F_minus}
\end{equation}
the second term being proportional to an incoming radial solution \(H_{lm}^-\) and the \(S\)-matrix. When there are no long-range potentials except for the possible central Coulomb monopole, the radial functions \(H_{lm}^\pm(r)\) are identical to the standard Coulomb-Hankel functions \(H_l^\pm(r) = G_l(r) \pm \mathrm{i}F_l(r)\)~\cite{DLMF}, which are eigensolutions of the one-particle Hamiltonian with Coulomb potential. For simplicity of notation we do not explicitly write out the charge and momentum dependences of these functions. In the high-energy limit the second term of Eq.~\eqref{eq:F_minus} becomes negligible and does not contribute to the phase of the wave function. The only relevant outgoing solutions have the asymptotics
\begin{equation}
    H_{lm}^+ \sim \exp [\mathrm{i}(kr + Z/k \cdot \ln 2kr - \pi l/2 + \sigma_l)] \,.
    \label{eq:Hasy}
\end{equation}
This asymptotic form must be preserved, including the overall phase, to ensure that the stationary photoionization state \eqref{eq:photo_wf} asymptotically converges to a Coulomb wave, or a plane wave in the absence of any residual charge, i.e. that it satisfies the physical boundary conditions.

In the case of a nonzero dipolar interaction between the residual molecule and the photoelectron use of Coulomb-Hankel functions for radial parts is an approximation, particularly in the absence of residual charge, because individual partial waves are coupled to a large distance due to the residual dipole. Even then, however, the outer region problem can be solved exactly~\cite{Burke,rageshkumar2022}. This is done by finding a set of radial functions \(H_{lm}^+\) that is an eigensolution of the one-electron Hamiltonian including the dipole-photoelectron interaction,
\begin{equation}
    \sum_{l'm'} \left[\left(-\frac{1}{2} \frac{\mathrm{d}^2}{\mathrm{d}r^2} + \frac{l(l + 1)}{2r^2} - \frac{Z}{r}\right)
    \delta_{ll'} \delta_{mm'} - \frac{\bm{D}_{ff} \cdot (\bm{n})_{lml'm'}}{r^2}\right] H_{l'm'}^+(r) = \frac{k^2}{2} H_{lm}^+(r) \,,
\end{equation}
where \((\bm{n})_{lml'm'} = \langle l m | \bm{r} | l' m' \rangle /r\).
The dipolar interaction can now be effectively absorbed into the centrifugal term,
\begin{equation}
    \sum_{l'm'} \left[\left(-\frac{1}{2} \frac{\mathrm{d}^2}{\mathrm{d}r^2} - \frac{Z}{r}\right)
    \delta_{ll'} \delta_{mm'} - \frac{A_{lml'm'}}{r^2}\right] H_{l'm'}^+(r) = \frac{k^2}{2} H_{lm}^+(r) \,,
    \label{eq:HAasy}
\end{equation}
where \(A_{lml'm'} = l(l+1)\delta_{ll'}\delta_{mm'} - 2\bm{D}\cdot(\bm{n})_{lml'm'}\) is a Hermitian matrix, which can be diagonalized. We denote the (real) eigenvalues of \(A\) by \(\lambda_j(\lambda_j + 1)\), where the generalized angular momentum \(\lambda_j\) is generally a complex number. We also denote by \(c_{lmj}\) the coefficients of the eigenvectors of \(A\). Then the solution of the coupled set of equations~\eqref{eq:HAasy} can be written as the linear combination
\begin{equation}
    H_{lm}^+(r) = \sum_j c_{lmj} H_{\lambda_j}^+(r) \,.
    \label{eq:Hexpansion}
\end{equation}
When this ansatz is substituted into Eq.~\eqref{eq:HAasy} and the resulting equation multiplied by the inverse of the matrix of eigenvectors \(c_{lmj}\), we recover equations for the standard Coulomb-Hankel functions
\begin{equation}
    \left(-\frac{1}{2} \frac{\mathrm{d}^2}{\mathrm{d}r^2} + \frac{\lambda_j(\lambda_j + 1)}{2r^2} - \frac{Z}{r}\right)
    H_{\lambda_j}^+(r) = \frac{k^2}{2} H_{\lambda_j}^+(r) \,.
    \label{eq:HAasy2}
\end{equation}
The asymptotic form of the functions \(H_{\lambda_j}^+\) is thus identical to~\eqref{eq:Hasy} with \(l\) replaced by \(\lambda_j\), which upon substitution into Eq.~\eqref{eq:Hexpansion} leads to the overall asymptotic formula for the outer solutions
\begin{equation}
    H_{lm}^+(r) \sim \exp [\mathrm{i}(kr + Z/k \cdot \ln 2kr)]
    \sum_j c_{lmj} \exp [\mathrm{i}(- \pi \lambda_j/2 + \sigma_{\lambda_j})] \,.
    \label{eq:Hasy_complex}
\end{equation}

The solutions \(H_{lm}^+(r)\) form a perfectly valid basis for partial-wave expansion of the photoelectron wave function that inherently includes the long-range dipole interaction with the residual molecule. However, if we directly substituted them into Eq.~\eqref{eq:F_minus} the resulting photoionization wave function~\eqref{eq:photo_wf} would not satisfy the correct Coulomb wave asymptotic boundary conditions since the special function \eqref{eq:Hasy_complex} has a more complicated asymptotic phase. For this reason, the additional term needs to be removed by changing the normalization of \(F^{(-)}\)
\begin{equation}
    F_{npk,flm}^{(-)}(r) \rightarrow \frac{\exp [\mathrm{i}(- \pi l/2 + \sigma_l)]}{\sum_\lambda c_{lm\lambda} \exp [\mathrm{i}(- \pi \lambda/2 + \sigma_\lambda)]} F_{npk,flm}^{(-)}(r) \,.
\end{equation}
Clearly, this eliminates from \(H_{lm}^+\) all effects of the dipole coupling to give a plain Coulomb-Hankel function. The effects of the permanent dipole are preserved only in the \(H^-\)-dependent term, which is discarded in the asymptotic approximation but it enters, via the $S$-matrix, the calculation of the Wigner delay. As a consequence all effects of the dipolar potential are already included in the Wigner delay and there are no additional asymptotic delays coming from the interplay between the permanent dipole and the IR absorption in \RABITT{}: .

\begin{figure}[htbp]
    \centering
    \includegraphics[width=0.5\textwidth]{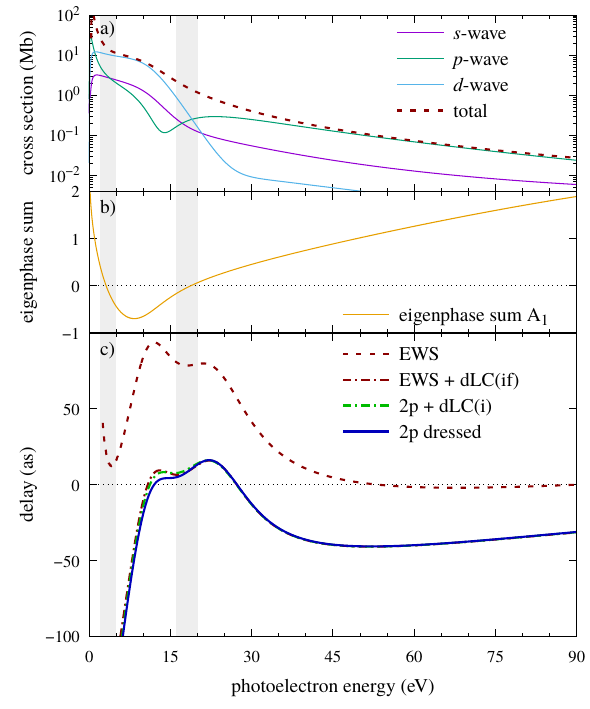}
    \caption{(a)~Total and partial wave emission-integrated cross section of one-photon ionization (detachment) of BeH\(^-\). (b)~Eigenphase sum of electron scattering on BeH in \(A_1\) symmetry. (c)~Wigner and \RABITT{} delays for photodetachment of BeH\(^-\) in the molecular frame, with photoemission along the molecular axis from the hydrogen end. Field polarization is parallel to the molecular axis.}
    \label{fig:BeH}
\end{figure}

We illustrate this conclusion for photodetachment of BeH\(^-\) in Fig.~\ref{fig:BeH}. This static-exchange calculation uses an $R$-matrix radius \(R_a = 200 a_0\), a 6-31G atom-centered basis, partial wave expansion up to \(\ell = 10\) and 400 radial $B$-splines of order 6. The large R-matrix radius is necessary to confine the diffuse initial bound state of the negative ion and to include the dominant part of the long-range dipole potential in the inner region, allowing us to treat the outer-region channel wave functions as effectively uncoupled. Convergence with respect to this parameter was tested by repeating the calculation with a radius of \(300 a_0\). As with LiH, we choose a very simple molecular model, in this case also to avoid excessive computational requirements imposed by the large inner region radius. However, a highly correlated calculation of the related electron scattering on neutral BeH is possible and has been performed with the molecular $R$-matrix method before~\cite{BeH-escat}. The calculated initial and final electronic dipole moments of the molecule were \(D_{ii} = -1.69\)~a.u. and \(D_{ff} = 1.64\)~a.u., respectively. In this case the addition of the dipole-laser coupling by means of Eq.~\eqref{eq:taudlc} to Wigner delay already results in complete agreement with the full two-photon calculation at high energies.

The calculated one- and two-photon delays are featureless at high photoelectron kinetic energies, but exhibit an enhancement region approximately between 5 and 30~eV. In this low-energy area, the magnitude of partial wave cross sections (Fig.~\ref{fig:BeH}a) varies strongly and the \(d\) wave dominant at low energies is replaced at high energies by dominance of the \(p\) wave, which recovers from its two-centre interference dip just below 15~eV. This large-scale structure in the EWS delay features two additional superimposed smaller dips in the delay highlighted by shaded areas in Fig.~\ref{fig:BeH} around photoelectron energies 3~eV and 18~eV. These areas coincide with the crossing of magnitudes of some partial wave cross sections in Fig.~\ref{fig:BeH}a. Interestingly, they also match energies of zeros in the calculated eigenphase sum for electron scattering on neutral BeH, Fig.~\ref{fig:BeH}b, in the only relevant symmetry \(A_1\) (irreducible representation of \(C_{2v}\)) accessible by the chosen parallel polarization. In electron scattering on atoms, vanishing of the scattering phase-shift is linked to Ramsauer-Townsend effect~\cite{Drake}, which can be interpreted as partial ``transparency'' of the target to the electron. Here we indeed observe reduction of the Wigner time delay in the grey areas, suggesting that the molecule affects the photoelectron a little less than at the surrounding energies. The analogy is imperfect, though, because in the molecular case the partial waves are coupled by the non-sphericity of the potential and one cannot assign unambiguous phase shifts to individual partial waves. When the additional IR probe photon is applied, positions of these dips move a little, but the overall large-scale structure remains unaffected.

\section{Conclusion}\label{sect:conclusion}

In this article we presented a time-independent model that allows inclusion of the IR-field-driven dynamics in the initial state in two-color ionization and applied it to the analysis of \RABITT{}. The field-dependent energy of a polar molecule translates to a superposition of time-independent dressed states. We combined this theory with the stationary molecular $R$-matrix approach for multi-photon transitions~\cite{multiphoton,RM-RABITT}.

Taking LiH as an example of a highly polar molecule, we demonstrated that the model agrees well with a nonperturbative fully time-dependent simulation and that the initial-state dynamics is indeed the missing piece in the second-order perturbation approach to \RABITT{}.

In the context of \RABITT{} we re-derived the approximate asymptotic formula for the dipole-laser coupling time delay known from the field of attosecond streaking~\cite{Pazourek,gebauer2019}. It was also shown that this formula is tightly related to the widely used molecular asymptotic theory of molecular \RABITT{} delays~\cite{BaykushevaWorner,RM-RABITT}. For this reason the formula gives accurate results at higher energies, but close to the ionization threshold it does not reproduce the full theory completely. Furthermore, we have proved and illustrated for the case of photodetachment of BeH\(^-\) that in the high-energy limit, there is no additional \RABITT{} delay component caused by coupling of the IR laser with the dipolar electron-molecule interaction: instead all dipolar effects are already included in the field-free one-photon (Wigner) delay.

As a consequence of the separability of the dipole-laser coupling at higher energies, the isotropic delays for a randomly oriented sample of molecules vanish at high energies. However, at low energies, even the orientation-averaged time delays still bear an imprint of the dynamics of the initial state. Nevertheless, we expect the effect to be very strong in the molecular frame over a wide range of energies.

\section{Acknowledgements}

JB and ZM acknowledge support of the Czech Science Foundation as the project GA CR 20-15548Y and support of the PRIMUS program of Charles University as the project PRIMUS/20/SCI/003. Computational resources were provided by the e-INFRA CZ project (ID:90254), supported by the Ministry of Education, Youth and Sports of the Czech Republic.

\bibliography{bibliography}

\end{document}